\documentclass[copyright,creativecommons,noderivs,noncommercial]{eptcs}
\providecommand{\event}{Gabriel Ciobanu (Ed.): Membrane Computing and\\ Biologically Inspired Process Calculi 2009} 
\providecommand{\volume}{15}   
\providecommand{\anno}{2009}   
\providecommand{\firstpage}{1} 
\providecommand{\eid}{1}       

\usepackage{amsmath,amsfonts,amsthm,amssymb}
\usepackage{latexsym,stmaryrd}
\usepackage{enumitem}
\usepackage{tikz}
\usetikzlibrary{arrows,calc}

\newcommand{\cat}[1]{{\normalfont\textsc{#1}}}

\newcommand{\defeq}{\triangleq}
\newcommand{\iffdef}{\stackrel{\triangle}{\iff}}
\newcommand{\cell}[2]{\ensuremath \Lbag #1 \lbag #2 \rbag \Rbag}
\newcommand{\sem}{\fatsemi}
\newcommand{\scomp}{\ast}
\newcommand{\mcomp}{\star}
\newcommand{\pinch}{\ensuremath \mathsf{p}}
\newcommand{\fuse}{\ensuremath \mathsf{f}}
\newcommand{\copinch}{\ensuremath \mathsf{p}^{\perp}}
\newcommand{\cofuse}{\ensuremath \mathsf{f}^{\perp}}
\newcommand{\set}[1]{\{#1\}}
\newcommand{\grow}{\rhd}

\newcommand{\kcomp}{\mathbin{,}}
\newcommand{\rlts}[2]{\frac{\displaystyle #1}{\displaystyle #2}}
\newcommand{\rl}[2]{\displaystyle\frac{#1}{#2}}
\newcommand{\zero}{\mathbf{0}}
\newcommand{\nil}{\diamond}
\newcommand{\dsb}[1]{\llbracket #1 \rrbracket}
\newcommand{\mext}{\textrm{\sf m}^\textrm{\sf ext}}
\newcommand{\mcis}{\textrm{\sf m}^\textrm{\sf cys}}
\newcommand{\pcarry}{\mathsf{p}^\mathsf{c}}
\newcommand{\pmemb}{\mathsf{p}^\mathsf{m}}
\newcommand{\pdir}{\mathsf{p}^\mathsf{d}}
\newcommand{\fcell}{\mathsf{f}^\mathsf{c}}
\newcommand{\fmemb}{\mathsf{f}^\mathsf{m}}
\newcommand{\fdir}{\mathsf{f}^\mathsf{d}}
\newcommand{\act}{\ensuremath act}
\newcommand{\Act}{\ensuremath \textsc{Act}}
\newcommand{\ok}{\ensuremath \mathsf{ok}}
\newcommand{\restrict}[1]{\ensuremath {{\upharpoonright}_{#1}}}

\newcommand{\biobeta}{Bio$\beta$}

\newtheoremstyle{theorem}
{\topsep}
{\topsep}
{}
{}
{\bfseries}
{.}
{0.5em}
{\thmname{#1}\thmnumber{ #2}\thmnote{ (#3)}}%

\newtheoremstyle{definition}
{}
{}
{}
{}
{\bfseries}
{.}
{0.5em}
{\thmname{#1}\thmnumber{ #2}\thmnote{ (#3)}}%

\theoremstyle{theorem} 
\newtheorem{proposition}{Proposition}[section]

\newtheorem{theorem}{Theorem}[section]

\theoremstyle{definition}
\newtheorem{definition}{Definition}[section]

\theoremstyle{remark}

\title{A framework for protein and membrane interactions}
\author{Giorgio Bacci
\institute{University of Udine}
\email{giorgio.bacci@dimi.uniud.it}
\and
Davide Grohmann
\institute{University of Udine}
\email{grohmann@dimi.uniud.it}
\and
Marino Miculan
\institute{University of Udine}
\email{miculan@dimi.uniud.it}
}

\def\titlerunning{A framework for protein and membrane interactions}
\def\authorrunning{G.~Bacci \& D.~Grohmann \& M.~Miculan}

\begin{document}

\maketitle

\begin{abstract}
  

  We introduce the \biobeta\ Framework, a meta-model for both
  protein-level and membrane-level interactions of
  living cells.  This formalism aims to provide a formal setting where
  to encode, compare and merge models at different abstraction levels;
  in particular, higher-level (e.g. membrane) activities can be given
  a formal biological justification in terms of low-level (i.e.,
  protein) interactions.

  A \biobeta\ specification provides a \emph{protein signature}
  together a set of \emph{protein reactions}, in the spirit of the
  $\kappa$-calculus.  Moreover, the specification describes when a
  protein configuration triggers one of the only two membrane
  interaction allowed, that is ``pinch'' and ``fuse''.

  In this paper we define the syntax and semantics of \biobeta,
  analyse its properties, give it an interpretation as biobigraphical
  reactive systems, and discuss its expressivity by comparing with
  $\kappa$-calculus and modelling significant examples.


  Notably, \biobeta\ has been designed \emph{after} a bigraphical
  metamodel for the same purposes.  Hence, each instance of the
  calculus corresponds to a bigraphical reactive system, and vice
  versa (almost).  Therefore, we can inherith the rich theory of
  bigraphs, such as the automatic construction of labelled transition
  systems and behavioural congruences.


\end{abstract}

\section{Introduction}

Cardelli in \cite{cardelli05:amsb} has convincingly argued that the
various \emph{biochemical toolkits} identified by biologists can be
described as a hierarchy of \emph{abstract machines},
each of which can be modelled using methods and techniques from
concurrency theory.  Like other complex situations, it seems unlikely
to find a single notation covering all aspects of a whole organism;
instead, several models, often presented as process calculi, have been
proposed in literature, each focusing on specific aspects of the
biological system, at different levels of abstractions.  This arises
the problem of how to \emph{integrate} these models.  Indeed, these
machines operate in concert and are highly interdependent: ``to
understand the functioning of a cell, one must understand also how the
various machines interact'' \cite{cardelli05:amsb}.  To this end, we
need a general \emph{meta}model, that is, a \emph{framework}, where
these models (possibly at different abstraction levels) can be
encoded, and their interactions can be formally described. Eventually
``we will need a single notation in which to describe all machines, so
that a whole organism can be described'' \cite{cardelli05:amsb}.

As a step towards this long-term goal, in this paper we present
\biobeta, a framework calculus for dealing with both protein-level and
membrane-level interactions.  Many specific models, with protein
interactions and membrane trafficking, can be encoded in \biobeta\ by
providing a \emph{specification}.  Notably, different specifications
can be merged, which corresponds to ``put together'' different models
and systems, often at different levels of abstraction.  This allows to
investigate the interactions between these models, which otherwise
would be kept separate; for instance, this allows to foresee
\emph{emerging properties}, such as behaviours due to interactions of
different abstraction levels, and which cannot be observed within a
single machine model due to its intrinsic abstractions.

A fundamental design choice is that this framework has to be
\emph{biologically sound}, i.e., it must admit only systems and
reactions which are biologically meaningful, especially at lower level
machines (i.e. protein).  This is important because in this way,
encoding a given model in the framework provides automatically a
formal, biologically sound justification for the model (or
``implementation'') in terms of protein reactions and explains how its
membrane-level interactions are realised by protein machinery.  Also,
the representation of a specific model is less error-prone.

To this end, \biobeta\ has been designed after a bigraphical
metamodel, called \emph{biobigraphs} and \emph{biological bigraphical
  reactive systems (BioRS)} and presented in a companion paper
\cite{bgm:biobig}, for dealing with both protein-level and
membrane-level interactions.  Actually, \biobeta\ can be seen as a
syntactic formalism for representing precisely biobigraphical reactive
systems.  However, for sake of simplicity and lack of space, in this
paper we will not discuss the connection between \biobeta\ and
biobigraphs, which will be presented in forthcoming work.  For the
purposes of the present paper, it suffices to know that biobigraphs,
and hence \biobeta, are \emph{adequate} with respect to Danos and
Laneve's \emph{$\kappa$-calculus}, one of the most accepted formal
model of protein systems: we can describe \emph{all} and \emph{only}
protein configurations and interactions of the $\kappa$-calculus.  (Is
important to notice, however, that our methodology is general, and can
be applied to other formal protein models.)  Hence, the \biobeta\
Framework can be seen somehow as an extension of the
$\kappa$-calculus, adding biological compartments (that is membranes);
however, this is a consequence of the fact that the underlying model,
the biobigraphs, are adequate w.r.t.~the $\kappa$-calculus.

When membranes come into play one wants to deal also with membrane
transport, hence, we have to take into consideration reactions like
endocytosis, exocytosis, cellular fusion and vesiculation.  Actually,
as observed by biologists (see \cite{au:phago} and
\cite[Ch.~15]{alberts:ecb}), membrane interactions present always a
``preparation phase'', where some proteins (receptors and ligands)
interact to get in place, followed by the actual ``membrane
reconfiguration'' phase, which can be either a \emph{pinch} or a
\emph{fuse} \cite{cardelli08:tcs}. Therefore, biobigraphs, and
\biobeta, provide just two general rules for all membrane
interactions, whilst the preparation phase depends on the specific
proteins involved and hence left to the encoding of the specific
model.

Summarizing, for encoding a given model one has just to instantiate
this general schema by specifying (a) a \emph{protein signature}, that
is a set of abstract proteins; (b) a set of \emph{protein rules},
describing protein interactions ignoring compartments; (c) a set of
\emph{protein configurations} which trigger a mobility reaction. Then,
all other reactions (mobility, in particular) are automatically
provided by the framework, in a sound way.  Hence, modelers can focus
on protein-level aspects, leaving to the framework the burden to deal
with biological compartments and membrane transport.


A further motivation for our framework comes from the many general
results provided by bigraphs. We mention here only the construction of
compositional bisimilarities \cite{milner:ic06}, allowing to prove
that two systems are \emph{observational equivalent}, that is, they
can be exchanged in any organism without that the overall behaviour
will change. Also for this application it is important to restrict the
systems allowed by the framework to only those biologically meaningful.



\paragraph{Synopsis}
In Section~\ref{sec:biobeta} we present the \biobeta\ framework: its
syntax, a type system for characterizing well-formed (i.e.,
biologically meaningful) terms, and general operational semantics.  As
an example application of this framework, in Section~\ref{sec:memtraf}
we model the mechanism of membrane traffic.  In
Section~\ref{sec:kappa} we provide a formal connection between
$\kappa$-calculus and \biobeta.  Related work is discussed in
Section~\ref{sec:relwork}, and in Section~\ref{sec:concl} we draw some
conclusion and plan future work.

\section{The \biobeta\ Framework}\label{sec:biobeta}

In this section we present the \biobeta\ framework.  Although one
should not think of \biobeta\ necessarily as an extension of the
$\kappa$-calculus, its connections with that model are quite strong,
and hence it is convenient to reuse notations and concepts from
$\kappa$.

\subsection{Syntax}

The language of the \biobeta\ framework is parametric over a
\emph{protein signature} $(\mathcal{P} = \mathcal{P}_{p} \cup
\mathcal{P}_{ap},s: \mathcal{P} \to \mathbb{N})$, where $\mathcal{P}$
is a finite set of (abstract) proteins, partitioned into \emph{polar}
$\mathcal{P}_{p}$ and \emph{polar} $\mathcal{P}_{ap}$.  $\mathcal{N}$
is an enumerable set of edge names, and $s\colon \mathcal{P} \to
\mathbb{N}$ assigns to every protein the number of its \emph{domain
  sites}.

A \emph{(protein) interface} is a map from $\mathbb{N}$
to $\mathcal{N} + \set{h,v}$ (ranged over $\rho$, $\sigma$, $\ldots$).
Given an interface $\rho$ and a protein name $A \in \mathcal{P}$, a
site $(A,i)$ is \emph{visible} if $\rho(i) = v$, \emph{hidden} if
$\rho(i) = h$, and \emph{tied} if $\rho(i)\in \mathcal{N}$. A site is
\emph{free} if it is visible or hidden.  In the following, we will
write $\rho = 1 + \bar{2} + 3^x$ to mean $\rho(1) = v$, $\rho(2) = h$,
and $\rho(3) = x$.  Denote with $|\rho,x|$ the number of occurrences
of $x$ in $\rho$, that is, $|\rho,x| \defeq |\set{i\in s(\rho) \mid
  \rho(i)=x}|$.

The syntax of \biobeta\ over a given protein signature, consists of
\emph{systems}, which can be nested, and \emph{membranes}.  Polar
proteins can freely float at the level of systems, while apolar
proteins can be embedded into membranes.  Let $A_{t} \in
\mathcal{P}_{t}$, for $t \in \set{p, ap}$, $n \in \mathcal{N}$, and
$\rho \colon \mathbb{N} \to \mathcal{N} + \set{h,v}$ be a protein
interface.
\begin{align}
  P,Q & ::= \nil \mid A_{p}(\rho) \mid \cell{S}{P} \mid P \scomp Q \mid
            \nu n. P\mid \pinch_n \sem P \mid \fuse_n \sem \cell{S}{P} 
            \tag{Systems} \\
  S,T & ::= \zero \mid A_{ap}(\rho) \mid S \mcomp T \mid
            \copinch_n \sem S \mid \cofuse_n
            \tag{Membranes} 
\end{align}

Systems are polar solutions (soluble in aqueous environments), which
can be either the empty system $\nil$, or a polar protein
$A_{p}(\rho)$, or a compartment $\cell{S}{P}$, or a group of systems
$P \scomp Q$, or a system prefixed by a \emph{new} name constructor
$\nu n.P$.  Membranes are apolar solutions (soluble in oily
environments), which can be the empty membrane $\zero$, or an apolar
protein $A_{ap}(\rho)$, or a group of apolar solutions $S \mcomp
T$. Notice that membranes are not sequences of actions (as, e.g., in
Brane calculus \cite{cardelli04:bc}) but simply a collection of
proteins.  As usual the ``new'' operator is a binder: in $\nu n.P$,
$P$ is the scope of the binder $\nu n$.  Sharing of names on
interfaces represent protein domain-domain bonds. For instance
$(x)(A(1^{x}) \kcomp B(1+2^{x})))$ is a protein solution where there
is a bond between sites $(A, 1)$ and $(B, 2)$.


\begin{figure}[t]
  \centering
  \begin{tikzpicture}[scale=0.6]
    \draw (0,0) coordinate (center)
    (-4,2) coordinate (upleft)
    (3.5,2) coordinate (upright)
    (8,2) coordinate (downleft)
    (14.5,2) coordinate (downright);
    
    \draw[very thick, draw=gray!80!blue, fill=blue!15, even odd rule]
    (upleft)
    plot[smooth cycle,tension=0.7] 
    coordinates{
      +(-1.5,1) +(0.6,1.2)
      +(1.1,0.4) + (1.5,0.65)
      +(2.5,0.5) +(2.5,-0.5)
      +(1.5, -0.65) + (1.1,-0.4) 
      +(0.6,-1.2) +(-1.5, -1)}
    (upleft)
    plot[smooth cycle,tension=0.7] 
    coordinates{
      +(-1.3,0.8) +(0.4,1)
      +(1.1,0.2) + (1.6,0.45)
      +(2.3,0.3) +(2.3,-0.3)
      +(1.6, -0.45) + (1.1,-0.2) 
      +(0.4,-1) +(-1.3, -0.8)};
    \draw[-triangle 45, purple, ultra thick] (upleft)++(right:2.7)
    +(left:0.6) -- +(right:0.6);
    \draw (upleft)++(0.5,-1.75) node {\textsf{vesciculation}};
			
    \draw[very thick, draw=gray!80!blue, fill=blue!15, even odd rule]
    (upright)
    plot[smooth cycle,tension=0.7] 
    coordinates{
      +(-1.5,1) +(0.6,1.2)
      +(1.1,0.2) + (0,0.45)
      +(-0.75,0.3) +(-0.75,-0.3)
      +(0, -0.45) + (1.1,-0.2) 
      +(0.6,-1.2) +(-1.5, -1)}
    (upright)
    plot[smooth cycle,tension=0.7] 
    coordinates{
      +(-1.3,0.8) +(0.4,1)
      +(0.9,0.4) + (0,0.65)
      +(-0.95,0.5) +(-0.95,-0.5)
      +(0, -0.65) + (0.9,-0.4) 
      +(0.4,-1) +(-1.3, -0.8)};
    \draw[triangle 45-, purple, ultra thick] (upright)++(left:0.8)
    +(left:0.6) -- +(right:0.6);
    \draw (upright)++(0,-1.75) node {\textsf{endocitosys}};
    
    \draw[very thick, draw=gray!80!blue, fill=blue!15, even odd rule]
    (downright)
    plot[smooth cycle,tension=0.7] 
    coordinates{
      +(-1.5,1) +(1,1)
      +(1,-1) +(-1.5, -1)}
    (downright)
    plot[smooth cycle,tension=0.7] 
    coordinates{
      +(-1.3,0.8) +(0.4,1)
      +(0.9,0.4) +(-0.2,0.65)
      +(-0.8,0.4) +(-0.8,-0.4)
      +(-0.2, -0.65) + (0.9,-0.4) 
      +(0.4,-1) +(-1.3, -0.8)}
    (downright)
    plot[smooth cycle,tension=0.7] 
    coordinates{
      +(-0.4,0.4) +(0.8,0.2)
      +(0.8,-0.2) +(-0.4,-0.4)};
    \draw[-triangle 45, orange, ultra thick] (downright)++(right:1.5)
    +(left:1) -- +(right:0.6);
    \draw (downright)++(0,-1.75) node {\textsf{exocitosys}};
    
    \draw[very thick, draw=gray!80!blue, fill=blue!15, even odd rule]
    (downleft)
    plot[smooth cycle,tension=1] 
    coordinates{
      +(-1.5,1) +(0,1)+(0,1)
      +(1.6,1)
      +(1.6,-1) +(0,-1)
      +(0,-1) +(-1.5, -1)
    }
    (downleft)
    plot[smooth cycle,tension=0.7] 
    coordinates{
      +(-1.3,0.7) +(-0.2,0.5)
      +(-0.2,-0.5) +(-1.3, -0.7)}
    (downleft)
    plot[smooth cycle,tension=0.7] 
    coordinates{
      +(1.4,0.7) +(0.3,0.5)
      +(0.3,-0.5) +(1.4, -0.7)};
    \draw[-triangle 45, orange, ultra thick] (downleft)++(up:0.3)
    +(left:0.6) -- +(right:0.6);
    \draw[triangle 45-, orange, ultra thick] (downleft)++(down:0.3)
    +(left:0.6) -- +(right:0.6);
    \draw (downleft)++(0,-1.75) node {\textsf{fusion}};
    
    \draw ($(center)+(up:1.5)+(right:0.25)$) node {\LARGE$\mathsf{pinch}$};
    \draw ($(center)+(up:1.5)+(right:11.25)$) node {\LARGE$\mathsf{fuse}$};
  \end{tikzpicture}
  \caption{Vesiculation and endocytosis can be generalized by means of
    an inward and outward pinch, and fusion and exocytosis by a
    horizontal and vertical fuse, respectively.}
  \label{fig:pinchfuse}
\end{figure}
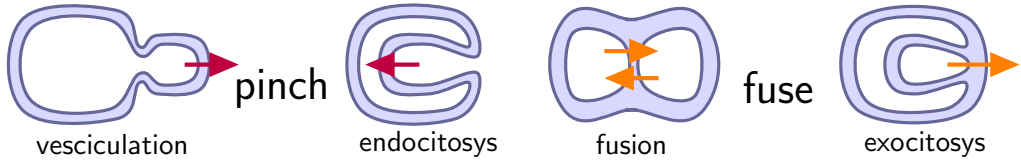

Finally, the special constructors \emph{pinch} $\pinch$ and
\emph{fuse} $\fuse$, together with their co-actions $\pinch^\perp,
\fuse^\perp$, to represent when a compartment reconfiguration takes
place.
These actions are inspired by the two basic membrane reconfiguration
processes, namely ``pinch'' and ``fuse''.  A membrane can be pinched
(either outward or inward, Figure~\ref{fig:pinchfuse} left), forming a
bubble that detaches producing a new compartment; or two membranes can
fuse, in either directions (Figure~\ref{fig:pinchfuse} right) mixing
contents.  Intuitively, $\pinch_n \sem P$ can be seen as a system $P$
waiting to complete, or \emph{commit}, a pinch interaction with
another system featuring the co-action $\pinch_n^\perp$; similarly for
$\fuse$. Thus, these actions are prefixes which freeze the
continuation system/membrane until the compartment reconfiguration is
completed. The subscripts pair-up corresponding actions and
co-actions.

Let us define the \emph{co-action} operator $(\cdot)^\perp$ mapping each mobility
action to its dual, and such that $((\cdot)^\perp)^\perp = id$; this
operator is extended to sets of actions as $S^\perp \defeq
\set{s^\perp \mid s\in S}$.  Moreover, for a set of names $X \subseteq
\mathcal{N}$, the \emph{restriction} of a set of actions over $X$ is
$S\restrict{X} \defeq \set{s_x\in S \mid x\in X}$.  In the following,
$P, Q, R, \ldots$ range over systems, $S, T, \ldots$ over membranes
and $K, L, H, \ldots$ over both systems or membranes.

The set of free names $fn(P)$ and $fn(S)$ are defined as usual,
covering also subscripted names on mobility actions (clearly
restriction on names binds also action names). In the following
$fn(K_1,\dots,K_n)$ stands for $\bigcup_{i=1}^n fn(K_i)$, for every
$K_i$ either a system or a membrane.

Let $\Act \defeq \set{\pinch_n,\copinch_n,\fuse_n,\cofuse_n \mid n \in
  \mathcal{N}}$ be the set of all mobility actions. We define the sets
of occurring actions $\act(P)$, $\act(S) \subseteq \Act$ on systems
and membranes, respectively, as follows:
\begin{gather*}
  \act(\nil) = 
  \act(A_{p}(\rho)) = \emptyset \qquad
  \act(\cell{S}{P}) = \act(S) \cup \act(P) \qquad
  \act(P \scomp Q) = \act(P) \cup \act(Q) \\
  \act(\zero) = 
  \act(A_{ap}(\rho)) = \emptyset \qquad
  \act(\nu n. P) = \act(P) \qquad
  \act(S \mcomp T) = \act(S) \cup \act(T)\\
  \act(\pinch_n \sem P) = \set{\pinch_n} \cup \act(P) \quad
  \act(\fuse_n \sem P) = \set{\fuse_n} \cup \act(P)  \quad
  \act(\copinch_n \sem S) = \set{\copinch_n} \cup \act(S) \quad
  \act(\cofuse_n) = \set{\cofuse_n}
\end{gather*}

\biobeta\ terms can be rearranged according to a \emph{structural
  equivalence} $\equiv$, defined to be the least equivalence closed
under $\alpha$-equivalence and satisfying the rules below:
\begin{gather*}
\begin{array}{c@{\qquad}c@{\qquad}c@{\qquad}c}
 \rl{}{P \scomp \nil \equiv P} &
 \rl{}{S \mcomp \zero \equiv S} &
 \rl{\mathsf{op} \in \set{\scomp,\mcomp}}
    {K \mathbin{\mathsf{op}} L \equiv L \mathbin{\mathsf{op}} K} &
 \rl{\mathsf{op} \in \set{\scomp,\mcomp}}
    {K \mathbin{\mathsf{op}} (L \mathbin{\mathsf{op}} H) \equiv
     (K \mathbin{\mathsf{op}} L) \mathbin{\mathsf{op}} H}
\end{array}
 \\[0.5ex]
\begin{array}{c@{\qquad}c@{\qquad}c@{\qquad}c}
 \rl{n \neq m}{\nu n. \nu m. P \equiv \nu m. \nu n. P} &
 \rl{}{\nu n. \nil \equiv \nil} &
 \rl{n \notin fn(Q)}{\nu n. (P \scomp Q) \equiv (\nu n. P) \scomp Q} &
 \rl{n \notin fn(S)}{\nu n.\cell{S}{P} \equiv \cell{S}{\nu n.P}}
\end{array}
\\[0.5ex]
\begin{array}{c@{\qquad}c@{\qquad}c@{\qquad}c@{\qquad}c}
 \rl{\mathsf{op} \in \set{\scomp,\mcomp} \quad K \equiv L}
    {K \mathbin{\mathsf{op}} H \equiv L \mathbin{\mathsf{op}} H} &
 \rl{P \equiv Q \quad S \equiv T}{\cell{S}{P} \equiv \cell{T}{Q}} &
 \rl{t \in \set{\pinch, \copinch, \fuse} \quad K \equiv L}
    {t_n \sem K \equiv t_n \sem L} &
\rl{t \in \set{\pinch, \fuse} \quad n \neq m}
   {\nu n. (t_m \sem P) \equiv t_m \sem \nu n. P}
\end{array}
\end{gather*}

\paragraph{Well-formedness}
As for other languages for graph-like structures, the syntax above is
too general, since many syntactically correct terms do not have a
clear biological meaning. In this subsection give an informal
description of the well-formedness conditions that terms must satisfy;
in the next subsection we will present a type system enforcing these
conditions.

Unsurprisingly, these conditions are similar to those of related
models, such as the $\kappa$-calculus, since they are motivated by
biochemical properties of proteins and membranes.

First, protein bonds can form only between exactly two active sites on
protein surfaces. Hence, terms like $(x)(A(1^{x}))$ or $(x)(A(1^{x})
\kcomp A(1^{x}) \kcomp B(1^{x}))$ have to be discarded.  This
requirement is extended to action names, because we want to pair-up
just a couple of actions at once. Furthermore, a pinch action can be
connected only to co-pinch, and similarly for fuse and
co-fuse\footnote{Actually, it suffice to disallow name sharing between
  actions and protein interfaces, but strengthening the condition will
  become useful in proving the subject reduction theorem.}.

Regarding membranes, we have to ensure the \emph{impermeability} of
membrane boundaries: complexations cannot take place through
membranes, hence protein bonds cannot cross them.  Thus, terms like
$A(1^{x}) \scomp \cell{\zero}{B(1^{x})}$ have to be ruled out.

Another (more technical) condition is that continuations after
mobility actions cannot have other actions. The intuition is that a
membrane engaged with a pinch or fuse interaction, has to complete it
before initiating another one.

We summarize these conditions in the follwing definition.
\begin{definition}[Well-formedness]\label{def:wf}
  A system $P$ is \emph{well-formed} if the following conditions hold:
  \setlist{noitemsep}
  \begin{description}[font=\itshape,labelindent=\labelsep,leftmargin=\parindent]
  \item[Graph-likeness:] free names occurs at most twice in $P$, and
    every binder ties either 0 or 2 occurrences;
  \item[Impermeability:] for any occurrence of a compartment
    $\cell{S}{Q}$ in $P$, $fn(Q) \subseteq fn(S)$ hold;
  \item[Action pairing:] for any occurrence of an action
    $t_{n}$ ($t \in \set{\pinch,\fuse}$), the name
    $n$ can appear only in the co-action $t_{n}^{\perp}$.
  \item[Action prefix:] for any occurrence of a prefix $t_n
    \sem Q$ $(t \in \set{\pinch, \copinch, \fuse})$ in $P$,
    $\act(Q) = \emptyset$ hold.
  \end{description}
\end{definition}

As expected, well-formedness is preserved by structural equivalence:
\begin{proposition}[Well-formedness is up to $\equiv$]
  Let $P$, $Q$ systems s.t.~$P \equiv Q$, $P$ is well-formed if it is
  $Q$.
\end{proposition}


\subsection{Type system}\label{sec:types}

In this subsection we present a type system for \biobeta\ terms, which
provides a formal procedure for checking well-formedness.  As usual in
type theory, terms are typed by judgements w.r.t.~some environment.

\begin{definition}[Judgement]
  A \emph{type} $\tau \subseteq \Act$ is a finite set of mobility actions.

  A \emph{judgement} for a \biobeta\ term $K$ (a system or a membrane)
  is of the form $\Gamma_1; \Gamma_2 \vdash K : \tau$, where the
  \emph{environment} $\Gamma_1;\Gamma_2$ is formed by two sets of
  names, such that $\Gamma_{1} \cap \Gamma_{2} = \emptyset$, each of
  which is written as a list of the form $x_0, \dots, x_n$.
\end{definition}
The two sets in the environment keep track of the free names in terms,
and the number of times each name occurs: in our case, if
$n\in\Gamma_1$ it occurs once, if $n\in \Gamma_2$ then it occurs
twice. Types keep track of mobility actions/co-actions occurring with
free names.

Moreover, let $\ok_{e}$ 
and $\ok_{t}$ 
be two predicates defined as follows:
\begin{equation*}
  \ok_{e}(\Gamma_1,\Gamma_2,\Delta_1, \Delta_2) \iffdef
  (\Gamma_1 \cup \Gamma_2) \cap (\Delta_1\cup \Delta_2) = \emptyset
  \qquad\qquad
  \ok_{t}(\Gamma,\tau,\sigma) \iffdef
    (\tau\restrict{\Gamma})^\perp = \sigma\restrict{\Gamma}
\end{equation*}
The first predicate checks if two environments share no names.  The
second predicate checks if the two types $\tau$, $\sigma$ pair-up
correctly, i.e., they create correct actions-coactions pairs using
names in $\Gamma$.

\begin{figure}[t]
  \centering
  \begin{gather*}
    (\textsf{empty}) \ 
    \rlts{\epsilon \in \set{\zero,\nil}}
         {\emptyset; \emptyset \vdash \epsilon : \emptyset}
         \qquad\qquad
    \rlts{A \in \mathcal{P} 
    \quad \forall x\in fn(\rho). \ |\rho,x| \leq 2}
         { \set{x\in fn(\rho) \mid |\rho,x| = 1};
           \set{x\in fn(\rho) \mid |\rho,x| = 2} \vdash A(\rho) : \emptyset
         }
         \ (\textsf{prot})
         \\[1ex]
         (\textsf{action}) \
    \rlts{ t\in \set{\pinch,\copinch,\fuse} \quad
           \Gamma_1; \Gamma_2 \vdash K : \emptyset \quad
           \act(K) = \emptyset
         }
         {\Gamma_1,x; \Gamma_2 \vdash t_x \sem K : \set{t_x}}
         \qquad\quad
    \rlts{ \Gamma_1; \Gamma_2 \vdash P : \tau \quad
           x \notin \Gamma_1 \quad
           \tau\restrict{\set{x}} = \emptyset
         }
         { \Gamma_1; \Gamma_2 \setminus \set{x}  \vdash
           \nu x. P : \tau
         }
         \ (\nu\textsf{-prot})
         \\[1ex]
         (\textsf{co-f}) \
    \rlts{}{x; \emptyset \vdash \cofuse_x : \set{\cofuse_x}}
         \qquad\qquad
    \rlts{ t \in \set{\pinch,\fuse} \quad
           \Gamma_1; \Gamma_2,x \vdash P : \tau \cup
           \set{t_x,t_x^{\perp}} \quad
           \set{t_x,t_x^{\perp}} \cap \tau = \emptyset
         }
         {\Gamma_1; \Gamma_2  \vdash \nu x. P : \tau}
         \ (\nu\textsf{-action})
         \\[1ex]
    (\textsf{par}) \
    \rlts{
      \begin{array}{c}
        \mathsf{op} \in \set{{\scomp},{\mcomp}} \quad
        \Gamma_1,\Gamma; \Gamma_2 \vdash K : \tau \quad
        \Delta_1,\Gamma; \Delta_2 \vdash L : \sigma \\[0.25ex]
        \ok_{e}(\Gamma_1,\Gamma_2,\Delta_1, \Delta_2) \quad
        \ok_{t}(\Gamma,\tau,\sigma)
      \end{array}
         }
         { \Gamma_1, \Delta_1;\Gamma_2, \Delta_2, \Gamma
           \vdash K \mathbin{\mathsf{op}} L : \tau \cup \sigma
         }
         \qquad
    \rlts{
      \begin{array}{c}
        \Gamma_1,\Gamma; \Gamma_2 \vdash S : \tau \quad
        \Gamma; \Delta_2 \vdash P : \sigma \\[0.25ex]
        \ok_{e}(\Gamma_1,\Gamma_2,\emptyset,\Delta_2) \quad
        \ok_{t}(\Gamma,\tau,\sigma)
      \end{array}
         }
         { \Gamma_1; \Gamma_2, \Delta_2, \Gamma
           \vdash \cell{S}{P} : \tau \cup \sigma
         }
         \ (\textsf{cell})
  \end{gather*}
  \caption{Type system for \biobeta\ terms.}
  \label{fig:typing}
\end{figure}

The \biobeta\ type system is shown in Figure~\ref{fig:typing}.
Intuitively, (\textsf{empty}) types empty systems and membranes, which
have no names and no actions.  (\textsf{prot}) deals with proteins
(possibly) having free names which fit in the environment, and no
actions.  The rule (\textsf{co-f}) types the singleton membrane
containing a co-fuse, which has a name $x$ of rank 1 and the set
containing itself as type.  (\textsf{action}) is for the remaining
actions: this rule checks that the subsystem contains no actions (thus
enforcing \textbf{action prefix} of Definition~\ref{def:wf}), adds a
\emph{fresh} name $x$ for the action, and create a type containing the
action itself.  The rules ($\nu$\textsf{-prot}) and
($\nu$\textsf{-action}) manage the restriction, in the first case
$\nu$ binds a name used by proteins, so the rule removes eventually
$x$ from names with rank 2 (remind that only names with 0 or 2
occurrences can be tied); in the second case there is a name $x$ of
rank 2 which connects a pair action-co-action, so the rule removes $x$
from $\Gamma_2$ as in previous case, and removes the actions from the
type (such actions have no free names anymore).  Last two rules are
for composing systems and membranes.  (\textsf{par}) puts side by side
two subsystem, checking that the global name rank is no greater then 2
by $\ok_e$ (enforcing \textbf{graph-likeness}) and that actions pairs
with respective co-actions by $\ok_t$ (\textbf{action pairing});
notice that common names ($\Gamma$) are promoted from rank 1 to 2.
For (\textsf{cell}) is similar: we build a cell from subsystems and,
beside the previous checkings, we impose that all names of rank 1 of
$P$ are contained in the ranked 1 names of $S$, i.e, $\Gamma_2$ is
empty; this enforces \textbf{impermeability}.

\begin{proposition}[Unicity of type]
  Let $K$ be a \biobeta\ system or a membrane. If $\Gamma_1;\Gamma_2
  \vdash K : \tau$ and $\Delta_1;\Delta_2 \vdash K : \sigma$, then
  $\Gamma_1 = \Delta_1$, $\Gamma_2 = \Delta_2$ and $\tau = \sigma$.
\end{proposition}

\begin{theorem}[Well-formedness]
  A \biobeta\ system $P$ is well-formed if and only if $\Gamma_{1};
  \Gamma_{2} \vdash P : \tau$, for some environment $\Gamma_{1}$,
  $\Gamma_{2}$ and type $\tau$.
\end{theorem}

\begin{proposition}[Subject congruence]
  Let $P$, $Q$ be two \biobeta\ systems and $S$, $T$ two \biobeta\
  membranes, such that $P \equiv Q$ and $S \equiv T$, then
  $\Gamma_{1}; \Gamma_{2} \vdash P : \tau$ if and only if $\Gamma_{1};
  \Gamma_{2} \vdash Q : \tau$ and $\Delta_{1}; \Delta_{2} \vdash S :
  \sigma$ if and only if $\Delta_{1}; \Delta_{2} \vdash T : \sigma$,
  for some $\Gamma_{1}$, $\Gamma_{2}$, $\Delta_{1}$, $\Delta_{2}$ and
  $\tau$, $\sigma$.
\end{proposition}

From now on, we consider only well-formed terms, except when it is
explicitly said.

\subsection{Semantics}\label{sec:semantics}

The operational semantics of a \biobeta\ system is given by a reaction
relation $\to$ on systems, defined by means of a rule system
parametrized over \emph{basic protein reactions} and \emph{mobility
  configurations}.

Defining a reaction semantics by means of a set of rules is the
simplest way for a biologist to look at protein reactions, since it
resembles usual chemical reactions.
Moreover, the rule-based modeling leaves biologists a great degree of
freedom in defining the behavior of a protein in solution. Such
modeling freedom is somewhat limited by biochemical constraints. One
of this was individuated and formalized by Danos and Laneve
in~\cite{dl:kappa04}: the \emph{causality} constraint.

Biochemical reactions are either complexations (i.e.~where low energy
bonds are formed on two complementary sites) or decomplexations,
possibly co-occurring with (de)activation of sites.  \emph{Causality}
does not allow simultaneous complexations \emph{and} decomplexations
on the same site. This constraint is assured by the notion of
\emph{(anti-)monotonicity}, which forces reactions not to decrease
(increase) the level of connections of a solution \cite{dl:kappa04}.
Moreover, protein reactions should be closed on all well-formed
membrane contexts, so that we are able to model reactions for proteins
located in different membranes and cells.

Let us now discuss about membrane level interactions, that is,
rearrangement of the membrane structure.  As observed by biologists,
these interactions result from complex nets of signaling pathways
which induce a mechanical reshaping of biological membranes, allowing
for example the formation of new vesicles, and hence new compartments.
However, as noticed before, actual membrane transformations are
limited and can be either a fuse or a pinch \cite{cardelli08:tcs};
what changes from situations to situations are the proteins involved
in the signaling pathway, leading to the actual fuse or pinch.  In
order to formalize this situation we will distinguish between the
effective membrane rearrangement, and what has trigger it.  More
precisely, a membrane reaction is split into two steps: a
``preparation phase'' where mobility actions ($\pinch$ or $\fuse$) are
introduced as prefixes in the system, and a ``commitment phase'' where
the structural reconfiguration of nested compartments is actually
performed. The commitment phase is formalized by rules defined once
and for all by the framework, corresponding to the pinch and fuse
processes of Figure~\ref{fig:pinchfuse}, but the preparation phase is
specified by the modeler. In other words, one has only to describe
\emph{when} these interaction take place by indicating which are the
suitable protein configurations that trigger a pinch or a fuse.  In
this way, each membrane interaction is given an explanation at the
protein-level.


\subsubsection{Protein reactions across multiple localities}
When dealing with protein reactions it is useful to think at complexes
as single entities which can be affected in all their protein
sub-units. An example of protein reaction we want to model is the
following:
\begin{equation*}
  D(1) \scomp A(1 + 2^{x}) \scomp \cell{B(1^{x} + 2^{y})}{C(1^{y} + \bar{2})} \to
  \nu z. \big ( 
  D(1^{z}) \scomp A(1^{z} + 2^{x}) \scomp \cell{B(1^{x} + 2^{y})}{C(1^{y} + 2)} 
  \big )
\end{equation*}
A complexation reaction between protein $D$ and the transmembrane
$A$$B$$C$-complex has effect not only in the formation of a new
protein bond between sites $(D,1)$ and $(A,1)$, but it toggles from
hide to visible site $(C, 2)$, which is not local to protein D
(promotor of the reaction). This is a very common situation in protein
signaling transduction, indeed transmembrane receptors (as
$ABC$-complex) propagate extracellular signals (as protein $D$) into
the intracellular environment according to this mechanism. 

Let us now consider the following reaction:
\begin{equation*}
  C(1^{y} + \bar{2}) \scomp \cell{B(1^{x} + 2^{y})}{A(1 + 2^{x}) \scomp D(1)} \to
  \nu z. \big ( 
  C(1^{y} + 2) \scomp \cell{B(1^{x} + 2^{y})}{A(1^{z} + 2^{x}) \scomp D(1^{z})} 
  \big )
\end{equation*}
This reaction differs from the previous one only for the direction in
which the ``signal'' is propagated. There is no biological reason to
distinguish between these two forms of reactions, indeed they are
exactly the same protein reaction.  In this sense, membranes have to
be taken into consideration only in the way they partition solutions
into distinct locations.

In order to define protein interactions in nested systems we introduce
\emph{(linear) contexts}. Contexts generalize the \biobeta\ syntax
introducing variables, which are sorted into two categories, systems
variables $X$ and membrane variables $Y$, used as ``holes'' for
systems and membrane, respectively.  Sequences $K_{1}, \dots, K_{n}$
are denoted by $\vec{K}$, and concatenation by $\vec{K}, \vec{K}' =
K_{1}, \dots, K_{n}, K_{1}', \dots, K_{m}'$.
\begin{align}
  C[\vec{X}; \vec{Y}], D[\vec{X}; \vec{Y}] & \mathrel{::=}
  	\nil \mid A_{p}(\rho)  \mid X \mid \pinch_n \sem P \mid
        \fuse_n \sem \cell{S}{P} \mid {} \notag \\
        & \phantom{{} \mathrel{::=} {}}
	C[\vec{X'}; \vec{Y'}] \scomp D[\vec{X''}; \vec{Y''}] \mid 
        \nu n. C[\vec{X}; \vec{Y}] \mid
	\cell{E[\vec{Y'}]}{C[\vec{X}; \vec{Y''}]}
   \tag{System contexts} \\
  E[\vec{Y}], F[\vec{Y}] & \mathrel{::=}
  	\zero \mid A_{ap}(\rho) \mid
	\copinch_n \sem S \mid \cofuse_n \mid
        E[\vec{Y'}] \mcomp F[\vec{Y''}] 
        \tag{Membrane contexts}
\end{align}
where $X \in \vec{X}$, $Y \in \vec{Y}$, and let $\pi,\pi'$ denote
permutations, $\vec{X} = \pi(\vec{X'},\vec{X''})$, $\vec{Y} =
\pi'(\vec{Y'},\vec{Y''})$.
Context application $C[\vec{P}; \vec{S}]$ is defined as expected, for
$\vec{P}$ and $\vec{S}$ two sequences of systems and membranes,
respectively.

As in the case of the $\kappa$-calculus, protein reactions can be of
two kinds: monotone and anti-monotone. Monotonicity is defined on a
growing relation ($\grow$) over \biobeta\ protein
solutions. Differently from $\kappa$-calculus, the notion of ``protein
solution'' in the \biobeta\ Framework have to take into account system
and membrane proteins, and the fact that they are in different
locations. Thus, we define protein solutions as sequences of system
groups and membrane groups of proteins, as follows:
\begin{definition}[Wide protein solutions]
  A \emph{group of system proteins} is a \biobeta\ system $P$ of the
  form $A_1(\rho_1) \scomp \ldots \scomp A_n(\rho_n)$, where $A_{i}
  \in \mathcal{P}_{p}$ ($1 \leq i \leq n$); similarly, a \emph{group
    of membrane proteins} is a \biobeta\ membrane $S$ of the form
  $B_1(\sigma_1) \mcomp \ldots \mcomp B_m(\sigma_m)$, where $B_{j} \in
  \mathcal{P}_{ap}$ ($1 \leq j \leq m$).

  A \emph{wide protein solution} is a couple of sequences of system
  groups and membrane groups of proteins, which can be restricted on
  some names $\vec{x}$, denoted as $\nu \vec{x}. \langle P_{1}, \dots,
  P_{k} \mid S_{1}, \dots, S_{h} \rangle$.
\end{definition}

Informally, a group of proteins is a \emph{local} protein solution,
that is, all proteins in that group reside in the same location. Wide
protein solutions are non-local and are formed as sequences of protein
groups of two kinds according with the syntactic separation of
membranes and systems.  Actually, wide protein solutions are not
\biobeta\ terms, but we use them as specifications of how complexes
are formed abstracting on localities.  Often, we will shorten $\nu
\vec{x}. \langle \vec{P} \mid \vec{S} \rangle$ as $\langle \vec{P}
\mid \vec{S} \rangle$ when $|\vec{x}| = 0$.  Free names are defined as
$fn(\nu x_{1}, \ldots x_{n}. \langle \vec{P} \mid \vec{S} \rangle) =
fn(\vec{P}, \vec{S}) \setminus \set{x_{1}, \ldots x_{n}}$.

We are interested in connected wide protein solutions, that is, wide
solutions where exists a path of bonds linking all the proteins in the
solution. Closed connected wide protein solutions are actually
\emph{protein complexes}.  Formally:
\begin{definition}[Connectedness]
 \label{def:connectedness}
  The \emph{connected} wide protein solutions are defined
  inductively as:
  \begin{gather*}
    \rl{}{\langle A(\rho) \mid {} \rangle}
    \qquad
    \rl{}{\langle {} \mid B(\sigma) \rangle}
    \qquad
    \rl{\nu \vec{x}. \langle \vec{P} \mid \vec{S} \rangle}
       {\nu x \vec{x}. \langle \vec{P} \mid \vec{S} \rangle}
    \qquad
    \rl{
      \begin{array}{c}
        \langle \vec{P} \mid \vec{S} \rangle  \qquad 
        \langle \vec{Q} \mid \vec{T} \rangle
        \\[0.3ex]
        fn(\langle \vec{P} \mid \vec{S} \rangle) \cap 
        fn(\langle \vec{Q} \mid \vec{T} \rangle) \neq \emptyset
      \end{array}
    }{\langle \vec{P},\vec{Q} \mid \vec{S}, \vec{T} \rangle}
    \\[0.3ex]
    \rl{
      \begin{array}{c}
        \langle \vec{P_{1}}, P, \vec{P_{2}} \mid \vec{S} \rangle \qquad 
        \langle Q \mid {} \rangle \\[0.3ex]
        fn(\langle \vec{P_{1}}, P, \vec{P_{2}} \mid \vec{S} \rangle) \cap 
        fn(\langle Q \mid {} \rangle) \neq \emptyset
      \end{array}
    }{\langle \vec{P_{1}}, P \scomp Q, \vec{P_{2}} \mid \vec{S} \rangle}
    \qquad
    \rl{
      \begin{array}{c}
        \langle \vec{P} \mid \vec{S_{1}}, S, \vec{S_{2}} \rangle \qquad 
        \langle {} \mid T \rangle \\[0.3ex]
        fn(\langle \vec{P} \mid \vec{S_{1}}, S, \vec{S_{2}} \rangle) \cap 
        fn( \langle {} \mid T \rangle) \neq \emptyset
      \end{array}
    }{\langle \vec{P} \mid \vec{S_{1}}, S \scomp T, \vec{S_{2}} \rangle}
  \end{gather*}
  A wide protein solution $\nu \vec{x}. \langle \vec{P} \mid \vec{S}
  \rangle$ is a \emph{wide complex} if it is connected and $fn(\nu
  \vec{x}. \langle \vec{P} \mid \vec{S} \rangle) = \emptyset$.
\end{definition}

\begin{definition}[Growing relation $\grow$]
  Let $\tilde{x}$ be a set of fresh names.  A growing relation $\grow$
  over interfaces is defined inductively as follows:
  \begin{equation*}
    \rlts{}{\tilde{x} \vdash \bar{i} \grow i} \quad \;
    \rlts{}{\tilde{x} \vdash i \grow \bar{i}} \quad \;
    \rlts{x\in \tilde{x}}{\tilde{x} \vdash i \grow i^x} \quad \;
    \rlts{\tilde{x}\cap fn(\rho)=\emptyset}{\tilde{x}\vdash\rho\grow\rho}
    \quad \;
    \rlts{\tilde{x}\vdash\rho\grow\sigma\quad\tilde{x}\vdash\rho'\grow\sigma'
    }{\tilde{x}\vdash \rho+\rho'\grow\sigma+\sigma'}
  \end{equation*}

  A growing relation $\grow$ over wide protein solutions is defined
  inductively as follows:
  \begin{gather*}
    \rl{}{\vec{x} \vdash \langle \nil \mid {} \rangle \grow
                         \langle \nil \mid {} \rangle}
    \qquad\quad
    \rl{}{\vec{x} \vdash \langle {} \mid \zero \rangle \grow
                         \langle {} \mid \zero \rangle}
    \qquad\quad
    \rl{ 
      \vec{x} \vdash \langle \vec{P} \mid \vec{S} \rangle \grow 
      \langle \vec{P'} \mid \vec{S'} \rangle
      \qquad
      \vec{x} \vdash \langle \vec{Q} \mid \vec{T} \rangle \grow 
      \langle \vec{Q'} \mid \vec{T'} \rangle
    }{\vec{x} \vdash \langle \vec{P},\vec{Q} \mid \vec{S},\vec{T} \rangle \grow 
      \langle \vec{P'},\vec{Q'} \mid \vec{S'},\vec{T'} \rangle
    }
    \\[1ex]
    \rl{\vec{x} \vdash \langle \vec{P},P \mid \vec{S} \rangle \grow 
      \langle \vec{Q},Q \mid \vec{T} \rangle
      \qquad
      \vec{x} \vdash \rho \grow \sigma
    }{\vec{x} \vdash \langle \vec{P},P\scomp A(\rho) \mid \vec{S} \rangle \grow 
      \langle \vec{Q},Q\scomp A(\sigma) \mid \vec{T} \rangle
    }
    \qquad\qquad
    \rl{\vec{x} \vdash \langle \vec{P} \mid \vec{S},S \rangle \grow 
      \langle \vec{Q} \mid \vec{T},T \rangle
      \qquad
      \vec{x} \vdash \rho \grow \sigma
    }{\vec{x} \vdash \langle \vec{P} \mid \vec{S},S\mcomp B(\rho) \rangle \grow 
      \langle \vec{Q} \mid \vec{T},T\mcomp B(\sigma) \rangle
    }
    \\[1ex]
    \rl{\vec{x} \vdash \langle \vec{P} \mid \vec{S} \rangle \grow 
      \langle \vec{Q},Q \mid \vec{T} \rangle
      \qquad
      fn(\rho) \subseteq \vec{x}
    }{\vec{x} \vdash \langle \vec{P} \mid \vec{S} \rangle \grow 
      \langle \vec{Q},Q\scomp A(\rho) \mid \vec{T} \rangle
    }
    \qquad\qquad
    \rl{\vec{x} \vdash \langle \vec{P} \mid \vec{S} \rangle \grow 
      \langle \vec{Q} \mid \vec{T},T \rangle
      \qquad
      fn(\sigma) \subseteq \vec{x}
    }{\vec{x} \vdash \langle \vec{P} \mid \vec{S} \rangle \grow 
      \langle \vec{Q} \mid \vec{T},T\mcomp B(\sigma) \rangle
    }
\end{gather*}
\end{definition}

Here, growing means possibly creating ``new'' bonds and/or proteins,
and toggling sites from visible to hide or vice versa.  Note that new
bonds use names from $\vec{x}$, that is, the following proposition
holds:
\begin{proposition}
  If $\vec{x} \vdash \langle \vec{P} \mid \vec{S} \rangle \grow
  \langle \vec{Q} \mid \vec{T} \rangle$, then $fn(\langle \vec{Q} \mid
  \vec{T} \rangle) = fn(\langle \vec{P} \mid \vec{S} \rangle) \cup
  \vec{x}$ and $\vec{x}\cap fn(\langle \vec{P} \mid \vec{S} \rangle) =
  \emptyset$.
\end{proposition}

On the concept of growing we define monotonicity, which is very
important to guarantee that the protein reactions respect the
principle of \emph{causality}.
\begin{definition}[Monotonicity and Anti-monotonicity]
  Let $\langle \vec{P} \mid \vec{S} \rangle$ and $\langle \vec{Q} \mid
  \vec{T} \rangle$ two wide protein solutions, we say that the pair
  $(\langle \vec{P} \mid \vec{S} \rangle, \nu \vec{x}. \langle \vec{Q}
  \mid \vec{T} \rangle)$ is \emph{monotone} if $\vec{x} \vdash \langle
  \vec{P} \mid \vec{S} \rangle \grow \langle \vec{Q} \mid \vec{T}
  \rangle$ and $\nu \vec{x}. \langle \vec{Q} \mid \vec{T} \rangle$ is
  connected.  $(\nu \vec{x}. \langle \vec{Q} \mid \vec{T} \rangle,
  \langle \vec{P} \mid \vec{S} \rangle)$ is \emph{anti-monotone} if
  $(\langle \vec{P} \mid \vec{S} \rangle, \nu \vec{x}. \langle \vec{Q}
  \mid \vec{T} \rangle)$ is monotone.
\end{definition}

We can now define what is a \emph{protein reaction specification}, and
the induced protein reactions.
\begin{definition}[Protein reaction specifications]
  A \emph{protein reaction specification} is a set $\mathcal{R}$ of
  monotone and/or anti-monotone pairs of wide protein solutions.

  Given a protein reaction specification $\mathcal{R}$, the
  corresponding \emph{protein reactions} are defined as follows:
  \begin{equation*}
    \rl{(\langle \vec{P} \mid \vec{S} \rangle, 
      \nu \vec{x}. \langle \vec{Q} \mid \vec{T} \rangle) \in \mathcal{R}
    }{C[\vec{P}; \vec{S}] \to \nu \vec{x}. C[\vec{Q}; \vec{T}]} \; \text{(mon)}
    \qquad\qquad\qquad
    \rl{(\nu \vec{x}. \langle \vec{Q} \mid \vec{T} \rangle,
      \langle \vec{P} \mid \vec{S} \rangle) \in \mathcal{R}
    }{\nu \vec{x}. C[\vec{Q}; \vec{T}] \to
                   C[\vec{P}; \vec{S}]} \; \text{(anti-mon)}
  \end{equation*}
  where $C[\vec{X}; \vec{Y}]$ is a context such that $\nu
  \vec{x}. C[\vec{Q}; \vec{T}]$ is well-formed (i.e., well-typed).
\end{definition}
In virtue of the generality of context $C[\vec{X}; \vec{Y}]$, we can
adopt a simplified notation for reactions: $\langle \vec{P} \mid
\vec{S} \rangle \to \nu \vec{x}. \langle \vec{Q} \mid \vec{T} \rangle$
for monotone reactions and, similarly, $\nu \vec{x}. \langle \vec{Q}
\mid \vec{T} \rangle \to \langle \vec{P} \mid \vec{S} \rangle$ for
anti-monotone ones.

\subsubsection{Membrane transport}

Membrane transformations are sorted into two groups: pinch ($\pinch$)
and fuse ($\fuse$). They occur only with the interaction of an action
$t_{n}$ and a co-action $t^{\perp}_{n}$
($t\in \set{\pinch, \fuse}$). Pinch and fuse can be applied
with different orientations, depending on where actions constructs are
placed.  Formally, the \emph{commitment rule} are
\begin{align}
  \pinch_n \sem P \scomp \cell{\copinch_n \sem S \mcomp T}{Q} & \to
  \cell{T}{\cell{S}{P} \scomp Q} \tag{pinch-in}\label{pinch-in} \\
  \cell{\copinch_n \sem S \mcomp T}{\pinch_n \sem P \scomp Q} & \to
  \cell{S}{P} \scomp \cell{T}{Q} \tag{pinch-out}\label{pinch-out} \\
  \fuse_n \sem \cell{S}{P} \scomp \cell{\cofuse_n \mcomp T}{Q} & \to
  \cell{S\mcomp T}{P \scomp Q} \tag{fuse-hor}\label{fuse-hor} \\
  \cell{\cofuse_n \mcomp T}{\fuse_n \sem \cell{S}{P} \scomp Q} & \to
  P \scomp \cell{S\mcomp T}{Q} \tag{fuse-ver}\label{fuse-ver}
\end{align}
We can pinch \emph{inward} or \emph{outward} a sub-system depending
on the direction where $\pinch$ and $\copinch$ are placed relatively
to each other.  Note that the continuations of $\pinch$ and $\copinch$
are used to form the new compartment, either in (\ref{pinch-in}) or
(\ref{pinch-out}).  Fusions can be either \emph{horizontal}
(\ref{fuse-hor}) or \emph{vertical} (\ref{fuse-ver}) according to the
relative positions of compartments to be merged.  Notice that, even in
these reactions, $\fuse$ and $\cofuse$ continuations are used to
select which part of the system has to be merged.

Mobility reactions respect \emph{bitonality} in the sense of
\cite{cardelli04:bc, cardelli08:tcs}, in fact the parity of nesting of
$P$ and $Q$ is preserved in all these reactions, hence they preserve
the bitonal coloring of those subsystems.

In nature, membrane transport occurs when a certain protein
configuration, allowing the mechanical reshaping of the membrane, is
reached.  Thus, in order to specify when a pinch or a fuse takes
place, it is sufficient to specify which are the protein
configurations leading to the ``appearing'' of mobility controls.

%
\begin{definition}[Membrane reaction specifications]
  A \emph{pinching configuration} is a 5-tuple $(P,P',S',S',Q)\in \Pi
  \times \Pi \times \Sigma \times \Sigma \times \Pi$, where $\Pi$,
  $\Sigma$ are the sets of action-free systems and action-free
  membranes, respectively.
  A \emph{fusing configurations} is 5-tuple $(P,S,R,T,Q) \in \Pi
  \times \Sigma \times \Pi \times \Sigma \times \Pi$.

  A \emph{membrane reaction specification} is a pair
  $(\mathcal{R}_{\pinch}, \mathcal{R}_{\fuse})$ where
  $\mathcal{R}_{\pinch}$ is a set of pinching configurations, and
  $\mathcal{R}_{\fuse}$ is a set of fusing configurations
  Given a membrane reaction specification $(\mathcal{R}_{\pinch},
  \mathcal{R}_{\fuse})$, the corresponding \emph{mobility introduction
    reactions} are defined as follows:
  \begin{gather}
    \rl{ (P,P', S, S', Q) \in \mathcal{R}_{\pinch} 
      \qquad fn(P) \subseteq fn(S) \qquad fn(S) \subseteq fn(P,Q)
      \qquad n \text{ fresh} }{
      P \scomp P' \scomp \cell{S \mcomp S' \mcomp T}{Q \scomp R} \to
      \nu n. (\pinch_{n} \sem P \scomp P' \scomp 
      \cell{\copinch_{n} \sem S \mcomp S' \mcomp T}{Q \scomp R})
    } \tag{intro-$\pinch$-in}
    \\[1ex]
    \rl{ (P,P', S, S', Q) \in \mathcal{R}_{\pinch} 
    \qquad fn(P) \subseteq fn(S) \qquad fn(S) \subseteq fn(P,Q)
    \qquad n \text{ fresh}}{
      Q \scomp \cell{S \mcomp S' \mcomp T}{ P \scomp P' \scomp R} \to
      \nu n. (Q \scomp \cell{\copinch_{n} \sem S \mcomp S' \mcomp T}{
        \pinch_{n} \sem P \scomp P' \scomp R})
    } \tag{intro-$\pinch$-out}
    \\[1ex]
    \rl{ (P, S, R, T, Q) \in \mathcal{R}_{\fuse} \qquad n \text{ fresh} }{
      \cell{S \mcomp S'}{P \scomp P'} \scomp R \scomp \cell{T \mcomp T'}{Q \scomp Q'} 
      \to
      \nu n. (\fuse_{n} \sem \cell{S \mcomp S'}{P \scomp P'} \scomp R \scomp 
      \cell{\cofuse \mcomp T \mcomp T'}{Q \scomp Q'})
    } \tag{intro-$\fuse$-hor}
    \\[1ex]
    \rl{ (P, S, R, T, Q) \in \mathcal{R}_{\fuse} \qquad n \text{ fresh}}{
      \cell{ T \mcomp T' }{ R \scomp \cell{ S \mcomp S' }{ P \scomp P' } \scomp Q \scomp Q'}
      \to
      \nu n. (\cell{\cofuse_{n} \mcomp T \mcomp T' }{ R \scomp 
        \fuse_{n} \sem \cell{ S \mcomp S' }{ P \scomp P' } \scomp Q  \scomp Q'})
    } \tag{intro-$\fuse$-ver}
  \end{gather}
\end{definition}

Intuitively, the 5-tuples in $\mathcal{R}_{\pinch}$ and
$\mathcal{R}_{\fuse}$ describe the systems configurations which
trigger a mobility action.  Notice that in the two pinch introduction
rules, the first two pairs of terms $(P, P')$ and $(S, S')$ in the
tuple correspond to two splits of sub-terms in the left hand side of
the reaction rule: this is for distinguishing the part that is
actually pinched ($P$ and $S$) from the part that just helps
triggering reaction but do not change its position ($P'$ and $S'$,
which may be enzymes). The same holds for fuse introduction rules,
where $P$ and $S$ will be fused with the compartment where $T$ and $Q$
occur, whereas $R$ just helps the fusion but will not change its
position.  In practice, actions are used for ``freezing'' the state of
the subsystem involved in the reconfiguration, until the
reconfiguration actually takes place.

Informally, $\mathcal{R}_{\pinch}$ and $\mathcal{R}_{\fuse}$ provide a
template of suitable conditions, cause of which membrane
reconfiguration can happen.  Such conditions are to be defined by the
modeler, who knows which are the requisites to make possible a pinch
(either for vesiculation or endocytosis) or a fusion (either for
fusion or endocytosis). This way of describing a model in the
\biobeta\ framework gives biologists (in general, modelers) the
possibility to justify membrane reconfigurations at the protein-level,
providing only the essential characteristics that a sub-system must
satisfy to perform a mobility reaction.

\medskip

Summing up, to define a \biobeta\ model, users have to provide the
protein signature, the set of protein rules $\mathcal{R}$, ignoring
compartments, and two sets of configurations $\mathcal{R}_{\pinch}$
and $\mathcal{R}_{\fuse}$ describing the situations for which a mobility
reaction can occur. All other reactions are automatically provided by
the framework.  Hence, modelers can focus on protein-level aspects,
leaving the framework to deal with biological compartments and
membrane transport.

\begin{definition}[\biobeta\ specification]\label{def:specif}
  A \biobeta\ specification is a quadruple $(\mathcal{P}, \mathcal{R},
  \mathcal{R}_{\pinch}, \mathcal{R}_{\fuse})$ where $\mathcal{P}$ is a
  protein signature, $\mathcal{R}$ is a protein reaction speciﬁcation,
  and $(\mathcal{R}_{\pinch}, \mathcal{R}_{\fuse})$ is a membrane
  reaction speciﬁcation.
\end{definition}

\begin{definition}[\biobeta\ reactive system]\label{def:reactsys}
  Let $(\mathcal{P}, \mathcal{R}, \mathcal{R}_{\pinch},
  \mathcal{R}_{\fuse})$ be a specification.
  The associated \emph{\biobeta\ reactive system} is a pair
  $(\Pi^{act}, \to)$, where $\Pi^{act}$ is the set of well-formed
  \biobeta\ systems over $\mathcal{P}$, and $\to$, called the
  \emph{reaction relation}, is the least binary relation over
  $\Pi^{act}$ containing all protein reactions over $\mathcal{R}$,
  mobility reactions \ref{pinch-in}, \ref{pinch-out}, \ref{fuse-hor},
  \ref{fuse-ver}, and all introduction reactions over
  $\mathcal{R}_{\pinch}$ and $\mathcal{R}_{\fuse}$, and it is closed
  under structural equivalence, system composition, restriction of
  names, and compartment nesting.
\end{definition}

Closure of $\to$ under system composition, restriction and
compartment constructs, allow to focus on the actual ``reacting
parts'' of the system.  Clearly, $\to$ is not closed under action
prefixes, because action prefixes ``freeze'' their continuation so
reactions cannot happen in them.

We can prove that for \biobeta\ reactive systems subject reduction
holds, that is, well-formedness for systems is preserved by reaction steps.
\begin{proposition}[Subject reduction]\label{prop:sr}
  Let $P$, $Q$ be two \biobeta\ systems.  If $\Gamma_{1}; \Gamma_{2}
  \vdash P : \tau$ and $P \to Q$, then $\Gamma_{1}; \Delta_{2} \vdash
  Q : \sigma$ where either $\Gamma_2 = \Delta_{2}$ and $\tau =
  \sigma$, or $\Gamma_2=\Delta_2,n$ and $\tau =
  \sigma+\{t_n,t_n^\perp\}$ for some $t\in\{\pinch,\fuse\}$.
\end{proposition}

\section{An example: Membrane Traffic}\label{sec:memtraf}

\begin{figure}[t]
  \centering
  (a)
  \includegraphics[width=0.45\textwidth]{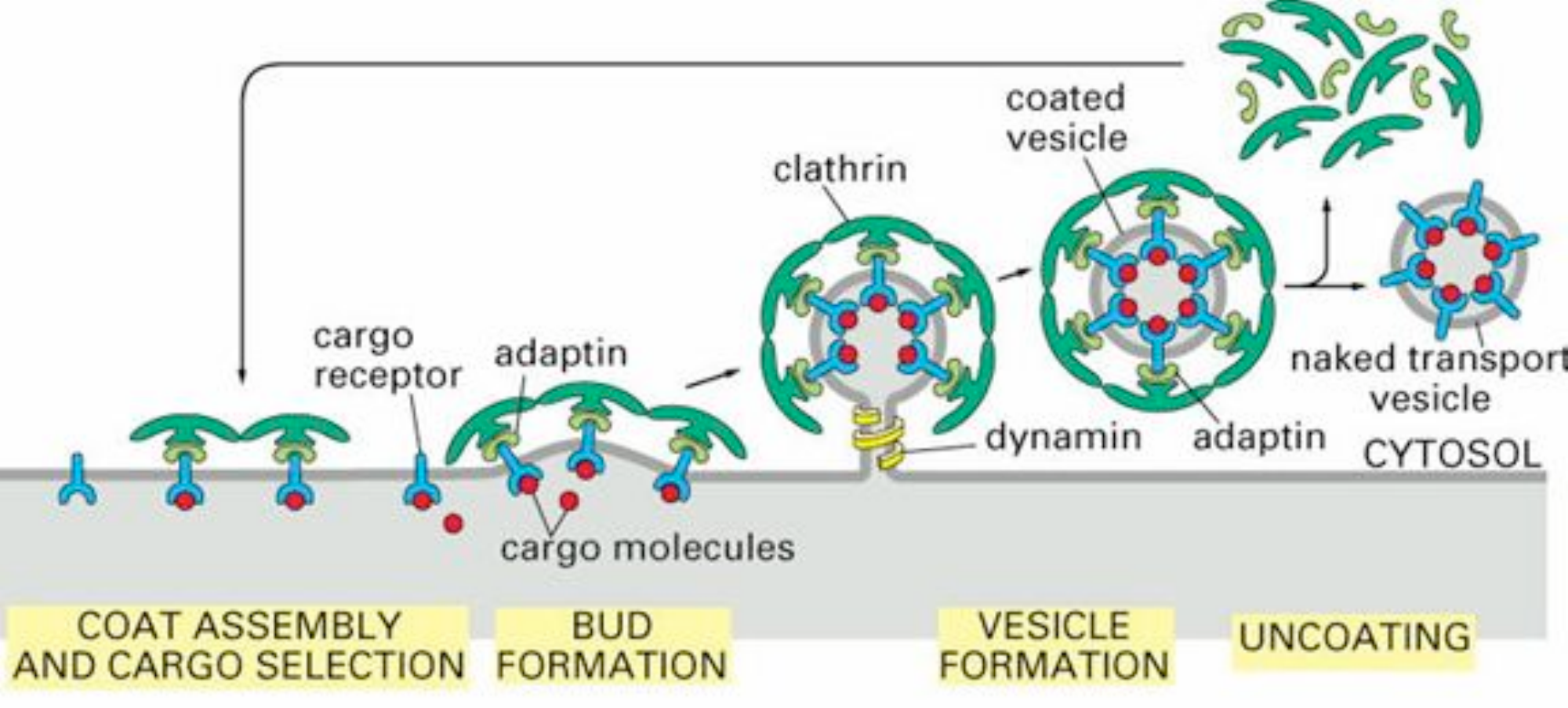}
  (b)
  \includegraphics[width=0.45\textwidth]{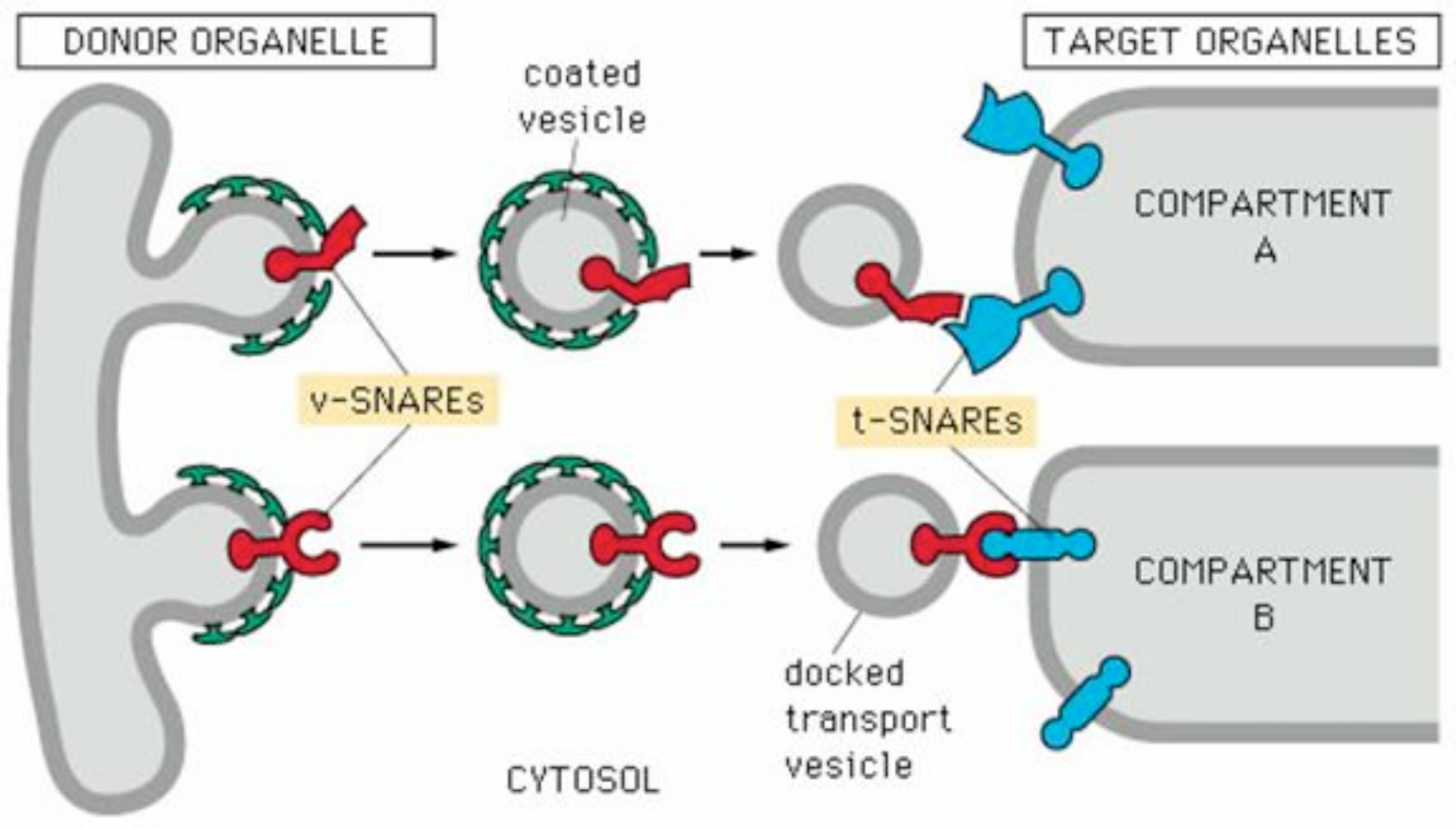}
  \caption{(a) Vesicle formation, (b) Vesicle docking. 
  (Pictures taken from \cite{alberts:ecb})}
  \label{fig:vtss}
\end{figure}

In this section we give a formal Bio$\beta$ description of how membrane
traffic works. Here we take two important processes into account, which are
fundamental in many transport mechanisms in cells: vesicle formation and 
vesicle docking.
Vesicles form by budding from membranes. Each bud has distinctive coat 
protein on cytosol surface, which shapes the membrane to form a bubble. 
The bud captures the correct molecules for outward transport by a selective
bond with a cargo receptor attached through the double layer to the coat protein
complex, which is lost after the budding completes (see Figure~\ref{fig:vtss}(a)). 
Vesicle contents are released in the target membrane by fusing vesicle with it. 

To ensure that membrane traffic proceeds in an orderly way, 
transport vesicles must be highly selective in recognizing the
correct target membrane with which to fuse. 
Because of the diversity of membrane systems, a vesicle is
likely to encounter many potential target membranes before it finds the correct 
one. Specificity in targeting is ensured displaying protein surface markers 
that identify them according to their origin and type of cargo 
(\emph{SNAREs}). Vesicles are marked by v-SNAREs 
(vector SNAREs) which selectively bind with some corresponding t-SNAREs 
(target SNAREs) located on target membrane (see Figure~\ref{fig:vtss}(b)).
SNARE proteins have a central role both in providing specificity and 
in catalyzing the fusion of vesicles with the target membrane.

Let $\mathcal{P}_{p} = \set{R_{i}^{e}, R_{i}^{c}, C_{i}, Ad_{i}, Cl, 
Sn_{i}^{e}, Sn_{i}^{c}, tSn_{i}^{e}, tSn_{i}^{c}}$ and 
$\mathcal{P}_{ap} = \set{ R_{i}^{m}, Sn_{i}^{m}, tSn_{i}^{m} }$, for $i \in \set{0,1}$.
Transmembrane proteins are represented by three sub-units in order to specify
the direction they exhibit with respect to membranes into which are embedded.
We denote with $R_{i}^{c}(1+2)$, $R_{i}^{m}(1+2)$, $R_{i}^{e}(1+2)$ 
the three cargo receptor sub-units, respectively for cytosolic, trans-membrane, 
and extra-cytosolic; SNAREs ($Sn^{*}_{i}$ and $tSn^{*}_{i}$) are
defined in a similar way.
$C_{i}(1)$ denotes a cargo molecule, $Ad_{i}(1+2)$ adaptin, and $Cl(1)$
clathrin. Proteins are sub-scripted with an index $i \in \set{0,1}$ to distinguish
different families of proteins.

We describe the \biobeta\ model for membrane traffic giving the set of reaction
rules for proteins and introduction of mobility. Protein reactions are defined as
follows:
\begin{align*}
  \langle
    C_i(1) \scomp R_i^e(1+2^x) , R_i^c(1^y+\bar{2}) \mid R_i^m(1^x+2^y)
  \rangle
  & \xrightarrow{\textsf{rec}}
  \nu z.
  \langle
    C_i(1^z) \scomp R_i^e(1^z+2^x) \kcomp R_i^c(1^y+2) \mid R_i^m(1^x+2^y)
  \rangle
  \\
  \langle
    R_i^c(1^x+2) \scomp Ad_i(1+\bar{2}) \mid
  \rangle
  & \xrightarrow{\textsf{adpt}}
  \nu y.
  \langle
    R_i^c(1^x+2^y) \scomp Ad_i(1^y+2) \mid
  \rangle
  \\
  \langle
    Ad_i(1^x+2) \scomp Cl(1) \mid
  \rangle
  & \xrightarrow{\textsf{coat}}
  \nu y.
  \langle
    Ad_i(1^x+2^y) \scomp Cl(1^y) \mid
  \rangle
  \\
  \nu y.
  \langle
    R_i^c(1^x+2^y) \scomp Ad_i(1^y+2^z) \scomp Cl(1^z) \mid
  \rangle
  & \xrightarrow{\textsf{uncoat}}
  \langle
    R_i^c(1^x+2) \scomp Ad_i(1+2^z) \scomp Cl(1^z) \mid
  \rangle
  \\
  \langle
    Sn_i^c(1^x+2) \scomp tSn_i^c(1+2^z) \mid
  \rangle
  & \xrightarrow{\textsf{snare}}
  \nu y.
    \langle
    Sn_i^c(1^x+2^y) \scomp tSn_i^c(1^y+2^z) \mid
  \rangle
\end{align*}
Intuitively, (\textsf{rec}) describes the formation of a bond between the cargo
molecule and the corresponding cargo receptor, which toggles to
visible the hidden site of the receptor cytosolic sub-unit. (\textsf{adpt}) and 
(\textsf{coat}) allow for the creation of the clathrin coat which leads to the formation
of the bud. (\textsf{uncoat}) frees the vesicle from the adaptin-clathrin coat.
Finally, (\textsf{snare}) deals with the snare-mediated docking of the naked vesicle,
by forming a bond between the v-SNARE and t-SNARE of the same family.
Figure~\ref{fig:vtss} gives an informal idea of how protein reactions work. 

Introduction rules are defined by the two sets of pinching and fusing 
configurations: $\mathcal{R}_{\pinch}$ and $\mathcal{R}_{\fuse}$. 

\begin{align*}
  \mathcal{R}_{\pinch} = \set{(P,P',S,S',Q)} \quad & \text{where} \quad
  \begin{aligned}
    P & = C_i(1^x) \scomp R_i^e(1^x+2^y) \scomp Sn_i^e(1^z) &
    P' & = \nil \\
    S & = R_i^m(1^y+2^w) \mcomp Sn_i^m(1^z+2^u) &
    S' & = \zero \\
    Q & = R_i^c(1^w+2^a) \scomp Ad_i(1^a+2^b) \scomp Cl(1^b) \scomp
    Sn_i^c(1^u+2)
  \end{aligned}
\end{align*}
The configuration $(P,P',S,S',Q)$ allows the vesicle to detach when
the adaptin-clathrin coat completely cover the forming bud. 
For sake of simplicity we consider coats of size one, i.e. with only one 
adaptin-clathrin complex (generalization to coat of size $n$ is straightforward).

\begin{align*}
  \mathcal{R}_{\fuse} = \set{(P,S,R,T,Q)} \quad & \text{where} \quad
  \begin{aligned}
    P & = C_i(1^x) \scomp R_i^e(1^x+2^y) \scomp Sn_i^e(1^a) &
    T & = tSn_i^m(1^d+2^e) \\
    S & = R_i^m(1^y+2^z) \mcomp Sn_i^m(1^a+2^b) &
    Q & = tSn_i^e(1^e) \\
    R & = R_i^c(1^z+2) \scomp Sn_i^c(1^b+2^c) \scomp tSn_i^c(1^c+2^d)
  \end{aligned}
\end{align*}
The configuration $(P,S,R,T,Q)$ describe the situation when the vesicle
is docked to the target membrane, hence the fusion can take place.

One possible initial configuration can be defined as follows: 
\begin{align*}
  Main & = DOrg \scomp Targ_0 \scomp Targ_1 \\
  DOrg & = Cl(1) \scomp Cl(1) \scomp
  Ad_0(1+\bar{2}) \scomp Ad_1(1+\bar{2}) \scomp
  R_0^c(1^{y_0}+\bar{2}) \scomp R_1^c(1^{y_1}+\bar{2}) \scomp
  Sn_0^c(1^{b_0}+2) \scomp Sn_1^c(1^{b_1}+2) \scomp {} \\
& \phantom{{}={}}
  \cell{
        R_0^m(1^{x_0}+2^{y_0}) \mcomp R_1^m(1^{x_0}+2^{y_1}) \mcomp
        Sn_0^m(1^{a_0}+2^{b_0}) \mcomp Sn_1^m(1^{a_1}+2^{b_1}) \\
        & \phantom{{}={} R_0^m(1^{x_0}+2^{y_0}) \mcomp
                   R_1^m(1^{x_0}+2^{y_1})} \,
       }
       {
        R_0^e(1+2^{x_0}) \scomp R_1^e(1+2^{x_1}) \scomp
        Sn_0^e(1^{a_0}) \scomp Sn_1^e(1^{a_1}) \scomp
        C_0(1) \scomp C_1(1)
       } \\
  Targ_0 & =
     tSn_0^c(1+2^{d_0}) \scomp \cell{tSn_0^m(1^{d_0}+2^{e_0})}{tSn_0^e(1^{e_0})} \\
  Targ_1 & =
     tSn_1^c(1+2^{d_0}) \scomp \cell{tSn_1^m(1^{d_1}+2^{e_1})}{tSn_1^e(1^{e_1})}
\end{align*}
In this case we have a single donor organelle which is able to pinch
out two vesicles (using two different type of SNAREs).  Such vesicle
can approach the corresponding docking membrane. 
In Figure~\ref{fig:vtss}(b) is depicted a possible run of computation.

\section{Comparing with $\kappa$-calculus}\label{sec:kappa}
In this section we give a formal connection between \biobeta\ and
Danos and Laneve's $\kappa$-calculus~\cite{dl:kappa04}.  This calculus
has been taken as the formal model protein interactions in the
definition of biobigraphs, the bigraphical model underlying the
\biobeta\ Framework.

The syntax of $\kappa$-solutions is the following:
\begin{equation*}
  S,T ::= \mathbf{0} \mid A(\rho) \mid S \kcomp T \mid (x)(S)
\end{equation*}
A $\kappa$-solution can be either the empty solution $\zero$, or a
protein $A(\rho)$, or a group of solutions $S \kcomp T$, or a solution
prefixed by name restriction $(x)(S)$ with $x \in \mathcal{N}$.

As usual in process calculi, a structural equivalence ($\equiv$) over
processes is introduced, which is the least structural containing
$\alpha$-equivalence and satisfying the abelian monoid on composition
and the scope extension and extrusion laws.  Moreover we define the
sets of free edge names $fn(\rho)$ and $fn(S)$, on interfaces and
solutions, defined in the common way, where the new name constructor
$(x)(S)$ is the only name binder.  We say a $\kappa$-solution $S$ is
\emph{closed} iff $fn(S) = \emptyset$.

Connectedness of a $\kappa$-solutions is particular a simper case of
the one seen before.
\begin{definition}[Connectedness and complexes]
  \emph{Connected $\kappa$-solutions} are defined inductively as:
  \begin{equation*}
    \rl{}{A(\rho)} \qquad
    \rl{S}{(x)(S)} \qquad 
    \rl{S\quad T\quad fn(S)\cap fn(T)\neq\emptyset}{S \kcomp T} \qquad
    \rl{S\quad S\equiv T}{T}
  \end{equation*}
   A $\kappa$-solution $S$ is a \emph{complex} if it is closed, graph-like, 
   and connected.
\end{definition}

The semantic of $\kappa$-calculus is defined by means of a transition
system built up by a user-defined set of monotone and anti-monotone
rules $\mathcal{R}_{\kappa}$.
\begin{definition}[Growing relation]
  A growing relation $\grow$ over $\kappa$-solutions is defined
  inductively as:
  \begin{align*}
    \rlts{}{\tilde{x} \vdash \mathbf{0}\grow\mathbf{0}} \qquad
    \rlts{\tilde{x}\vdash S\grow T\quad\tilde{x}\vdash\rho\grow\sigma
    }{\tilde{x}\vdash S\kcomp A(\rho)\grow T\kcomp A(\sigma)} \qquad
    \rlts{\tilde{x}\vdash S\grow T\quad fn(\sigma)\subseteq\tilde{x}
    }{\tilde{x}\vdash S \grow T\kcomp A(\sigma)}
  \end{align*}
\end{definition}

Again, monotonicity is a simplification of the one shown for \biobeta.
\begin{definition}[Monotone reactions]
  Let $L,R$ be two $\kappa$-solutions. $L \rightarrow (\tilde{x})R$ is
  a \emph{monotone reaction} if $\tilde{x}\vdash L\grow R$, where
  $L,(\tilde{x})R$ graph-like and $R$ connected.  $ (\tilde{x})L
  \rightarrow R$ is \emph{antimonotone} if $R \rightarrow
  (\tilde{x})L$ is monotone.
\end{definition}

Finally, we are able to characterize a protein transition system:
\begin{definition}[Protein transition systems]
  Given a set $\mathcal{R}_\kappa$ of monotone and antimonotone
  reactions, we define a \emph{protein transition system} (PTS) as a
  pair $\langle \mathcal{S},\leadsto\rangle$, such that $\mathcal{S}$
  is a set of $\kappa$-solutions, and $\leadsto$ is the least binary
  relation over $\mathcal{S}$ such that $\mathcal{R}_\kappa\subseteq
  {\leadsto}$, closed w.r.t.~$\equiv$, composition and name
  restriction.
\end{definition}

Now we are able to establish formally a connection between the
\biobeta\ framework and the $\kappa$-calculus, that is all protein
reactions that can be take place in our systems are justifiably by
means of protein reactions in a $\kappa$-solution.  In order to do
this, first we show a way to translate \biobeta\ systems and reactions
into $\kappa$ solutions and reactions ($\dsb{-}$). $\dsb{-}$ is
defined as follows:
\begin{gather*}
  \begin{aligned}
    \dsb{\nil} & = \zero \\    
    \dsb{\zero} & = \zero \\
    \dsb{\pinch_n \sem P} & = \dsb{P}
  \end{aligned}
  \qquad
  \begin{aligned}
    \dsb{A_p(\rho)} & = A_p(\rho) \\
    \dsb{A_{ap}(\rho)} & = A_{ap}(\rho) \\
    \dsb{\fuse_n \sem P} & = \dsb{P}
  \end{aligned}
  \qquad
  \begin{aligned}
    \dsb{P \scomp Q} & = \dsb{P} \kcomp \dsb{Q} \\
    \dsb{S \mcomp T} & = \dsb{S} \kcomp \dsb{T} \\
    \dsb{\copinch_n \sem S} & = \dsb{S}
  \end{aligned}
  \qquad
  \begin{aligned}
    \dsb{\cell{S}{P}} & = \dsb{S} \kcomp \dsb{P} \\
    \dsb{\nu n.P} & = (n) (\dsb{P}) \\
    \dsb{\cofuse_n} & = \zero
  \end{aligned}
\end{gather*}

Such encoding induces a type system over the $\kappa$-solution for
checking if the solutions are graph-like.  It can be given as follows:
\begin{gather*}
  (\textsf{zero}) \
  \frac{}{\emptyset ; \emptyset \vdash \zero}
  \qquad\qquad\qquad
  \frac{A\in \mathcal{P} \quad \forall x\in fn(\rho).|\rho,x| <2}
       {\set{x\in fn(\rho) \mid |\rho,x| = 1} ;
        \set{x\in fn(\rho) \mid |\rho,x| = 2} \vdash A(\rho)
       }
  \ (\textsf{prot})
  \\[0.5ex]
  (\textsf{res}) \
  \frac{\Gamma_1 ; \Gamma_2 \vdash S \quad x\notin \Gamma_1}
       {\Gamma_1 ; \Gamma_2 \setminus \set{x} \vdash (x) S}
  \qquad\qquad
  \frac{\Gamma_1,\Gamma ; \Gamma_2 \vdash S \quad
        \Delta_1, \Gamma ; \Delta_2 \vdash T \quad
        (\Gamma_1 \cup \Gamma_2) \cap (\Delta_1 \cup \Delta_2) = \emptyset
       }{\Gamma_1,\Delta_1;\Gamma_2,\Delta_2,\Gamma \vdash S \kcomp T}
  \ (\textsf{par})
\end{gather*}
The following properties stating a statical relation between \biobeta\
and $\kappa$ hold.
\begin{proposition}[Syntax]
  \begin{enumerate}
  \item For $S$ a $\kappa$-solution, if $\Gamma_1 ; \Gamma_2 \vdash
    S$ and $\Delta_1 ; \Delta_2 \vdash S$ then $\Gamma_1{=}\Delta_1$
    and $\Gamma_2 {=} \Delta_2$.
  \item For $S$ a $\kappa$-solution,  $S$ is graph-like iff
    $\Gamma_1 ; \Gamma_2 \vdash S$ for some $\Gamma_1,\Gamma_2$.
  \item For $S,T$ two $\kappa$-solutions,  if $S\equiv T$, then
    $\Gamma_1 ; \Gamma_2 \vdash S$ iff $\Gamma_1 ; \Gamma_2 \vdash T$.
  \item For $P$ a \biobeta\ system, if $\Gamma_1 ; \Gamma_2 \vdash P
    : \tau$, then $\Gamma_1 ; \Gamma_2 \vdash \dsb{P}$.
  \end{enumerate}
\end{proposition}

Finally, we can state and prove the following semantics/dynamic
relation between \biobeta\ and $\kappa$:
\begin{proposition}[Semantics]
  \begin{enumerate}
  \item For $S,T$ two $\kappa$-solutions, if $\Gamma_1 ; \Gamma_2
    \vdash S$ and $S \leadsto T$, then $\Gamma_1 ; \Gamma_2 \vdash T$.
  \item For a wide monotone protein reaction $(\langle
    \vec{P},\vec{S}\rangle, \nu \vec{x}.\langle
    \vec{P}',\vec{S}'\rangle)\in \mathcal{R}$, and for every
    $C[\vec{X},\vec{Y}]$ such that $C[\vec{P},\vec{S}]$ is
    well-formed, then $\dsb{C[\vec{P},\vec{S}]} \leadsto \dsb{\nu
      \vec{x}.C[\vec{P}',\vec{S}']}$.
  \item For a wide anti-monotone protein reaction $(\nu
    \vec{x}.\langle \vec{P},\vec{S}\rangle,\langle
    \vec{P}',\vec{S}'\rangle)\in \mathcal{R}$, for every
    $C[\vec{X},\vec{Y}]$ such that $\nu \vec{x}. C[\vec{P},\vec{S}]$
    is well-formed, then $\dsb{\nu \vec{x}. C[\vec{P},\vec{S}]}
    \leadsto \dsb{C[\vec{P}',\vec{S}']}$.
  \end{enumerate}
\end{proposition}
Intuitively, this proposition states that every protein transition
performed in the \biobeta\ framework is justified by a corresponding
transition in the $\kappa$-calculus.  The converse does not hold, due
to the compartment confinements: not all reactions of the translation
of a \biobeta\ system which are possible in $\kappa$-calculus,
correspond effectively to some wide (anti)monotone protein reactions
in the \biobeta\ framework.

\section{Related works}\label{sec:relwork}

In literature several other calculi, covering both protein reactions
and membrane interactions, have been proposed.  Two particularly
interesting cases are the \textsf{bio}$\kappa$-calculus
\cite{lt:biokappa} and the $\mathcal{C}$-calculus
\cite{ddk08:langcell}.  In this section we compare our framework with
these calculi analyzing analogies and differences of the approaches.

\paragraph{\textsf{bio}$\kappa$-calculus} can be seen as a version of the
$\kappa$-calculus extended with compartments. Compared to \biobeta,
the most remarkable similarities are at the syntactic level. Both
calculi have a compartment constructor which embeds proteins into
membranes, and similar well-formedness conditions to restrict to only
biological relevant processes.  Both calculi are sound at protein
level in the sense of the $\kappa$-calculus. In both cases
preservation of well-formedness during reactions is proved, although
\textsf{bio}$\kappa$-calculus does not have a formal type system as
\biobeta.

Several differences emerge at the semantic level, though.  An
important design choice distingushing \textsf{bio}$\kappa$ and
\biobeta\ concerns how the behaviour of a compartment is specified.
In \textsf{bio}$\kappa$, the mobility capabilites of a compartment are
indicated by its \emph{name}, which appears explicitly in the
rearrangement rules which can be applied.  The result of a membrane
interaction is one or more new membranes whose names have to be
specified in the rearrangement rule. This means that the rule defines
also the future (mobility) behaviour of the resulting systems; hence
it is up to the modeller to specify correctly the evolutions of names
of membranes.

On the other hand, in \biobeta\ compartments have no names: membranes
are just containers and their behaviour is completely determined by
the interactions of proteins floating in the aqueous solutions and
lipid bi-layers.  Therefore, a modeller has only to provide
protein-level rules and define the configurations which trigger a
mobility reaction. One does not have to declare how a membrane will
evolve (i.e., ``which is the next name'' in \textsf{bio}$\kappa$),
because the behaviour is defined by the rules of the framework.

Due to these radically different design choices, a formal comparison
at the semantic level between \textsf{bio}$\kappa$-calculus and
\biobeta\ is not easy.  Moreover, the \textsf{bio}$\kappa$ semantics
is given by means of an LTS, instead \biobeta\ has a reaction
(reduction) relation.  A possible future work is to derive an LTS with
associated congruential bisimilarity, taking advantage of the theory
of bigraphs.

\paragraph{$\mathcal{C}$-calculus}
Both $\mathcal{C}$-calculus and \biobeta\ frameworks are founded on
bigraphical reactive systems, which does not only confer a formal
graphical representation but provides also a powerful categorical
theory for automatically constructing labelled transition systems
(LTSs) whose bisimilarities are always congruences.  Even though
$\mathcal{C}$-calculus and \biobeta\ share common sources of
inspiration, they differs in many aspects, one of which is the way
compartment reconfiguration are performed.

$\mathcal{C}$-calculus adopts meta-biological ``gates'' to represent a
semi-fused intermediate state in membrane fusion or fission. Gates act
like channels, allowing protein diffusion between compartments, and it
does not allow for phagocytosis.  Moreover, one of the aim of
\biobeta\ framework is to characterize higher-level membrane
interactions with lower-level protein reactions, and to be adequate
with respect to the $\kappa$-calculus; a formal connection between
$\mathcal{C}$-calculus and $\kappa$-calculus has not been investigated
yet.  Finally, in the $\mathcal{C}$-calculus, mobility and protein
reactions are more intertwined, since membrane interactions are in
many steps: introduction of gates, some protein diffusions, and
deletion of gates. This mechanism does not provide an explicit,
logical separation between compartment reconfiguration and protein
reactions.  On the other hand, the \biobeta\ framework aims to an
explicit distinction between protein reactions and mobility actions,
as implemented by means of introduction rules, which freeze the
subsystems involved in the rearrangement until the structural
reconfiguration is actually performed.


\section{Conclusion}\label{sec:concl}

In this paper we have presented the \biobeta\ Framework, a language
for both (lower-level) protein and (higher-level) membrane
interactions of living cells.  A typed language has been given for
describing only biologically meaningful compartments and proteins.

In the \biobeta\ framework membrane reconfigurations are logically
different from protein reactions, but the connection between the two
aspects is formally specified.  In fact, membrane activities can be
motivated by a formal biological justiﬁcation in terms of protein
interactions.  Notably, protein interactions are adequate with respect
to the $\kappa$-calculus.

As an example application of this framework, we have modelled the
mechanism of membrane traffic, which consists in formation,
transportation and docking of vesicles.

\smallskip
\noindent\textit{Related and future work.}
Due to lack of space, we have omitted a formal connection between the
proposed framework and its inspiring bigraphical model.  Another
interesting aspect to investigate is the bisimilarity definable via
the IPO construction on bigraphs \cite{milner:ic06}, which is always a
congruence. This can be useful e.g.~in \emph{synthetic biology}, for
instance, to verify if a synthetic protein (or even a subsystem to be
implanted) behaves as the natural one.  We plan to describe such results
in a forthcoming work.

Another interesting investigation is to give a formal comparison of
\biobeta\ with the other two framework present in literature, such as
\textsf{bio}$\kappa$ and $\mathcal{C}$-calculus.  In particular, it
could be useful to understand the differences of
$\mathcal{C}$-calculus rule generators and our introduction rules,
even because they are both inspired by bigraphical wide reactions.

As a next step, a stochastic version of the \biobeta\ framework should
be analyzed for simulation purposes. One way could be trying to adapt
the extended Gillespie simulation algorithm
\cite{versari-busi:cmsb07}, that works for calculi with
multi-compartments.

\bibliographystyle{plain}
\bibliography{allbib}

\end{document}

\section{Encoding in BioBigraphs}

Encoding function from systems and membranes with actions in
biobigraphs:
\begin{alignat*}{2}
  \drb{\nil}_X^M & = \drb{\zero}_X^M = 1 \parallel X \\
  \drb{\nu n. Q}_X^M & = /n{:\,} \mathsf{h} \circ \drb{Q}_{X \uplus \set{n}}^M \\
  \drb{Q \scomp R}_X^M & = \drb{Q}_X^M \mid \drb{R}_X^M \\
  \drb{\pinch_n \sem Q \scomp R}_X^M & =
         \pcarry_n \circ \drb{Q}_X^\emptyset \mid \drb{R}_X^{M\uplus\set{\pinch_n}}
  & \quad & n \in X \\
  \drb{\pinch_n \sem Q \scomp R}_X^{M \uplus \set{\pinch_n}} & =
         \pcarry_n \circ \drb{Q}_X^\emptyset \mid \drb{R}_X^M
  & \quad & n \in X \\
  \drb{\fuse_n \sem \cell{S}{Q} \scomp R}_X^M & =
         \fcell_n \circ \drb{\cell{S}{Q}}_X^\emptyset \mid
         \drb{R}_X^{M\uplus \set{\fuse_n}}
  & \quad & n \in X \\
  \drb{\fuse_n \sem \cell{S}{Q} \scomp R}_X^{M\uplus \set{\fuse_n}} & =
         \fcell_n \circ \drb{\cell{S}{Q}}_X^\emptyset \mid \drb{R}_X^M
  & \quad & n \in X \\
  \drb{S \mcomp T}_X^M & = \drb{S}_X^M \mid \drb{T}_X^M \\
  \drb{\cell{S}{Q}}_X^M & =
  \mext \circ (\drb{S}_X^M \mid \mcis \circ \drb{Q}_X^M) \\
  \drb{\cell{\copinch_n \sem S \mcomp T}{Q}}_X^M & =
  /m{:\,} \mathsf{h} \circ (\pdir_m \mid \mext \circ
  (\pmemb_{n,m} \circ \drb{S}_X^\emptyset \mid \drb{T}_X^M
               \mid \mcis \circ \drb{Q}_X^{M\uplus \set{\pinch_n}}))
  & \quad & n\in X,\ m\notin X \\
  \drb{\cell{\copinch_n \sem S \mcomp T}{Q}}_X^{M\uplus \set{\pinch_n}} & =
  /m{:\,} \mathsf{h} \circ \mext \circ
    (\pmemb_{n,m} \circ \drb{S}_X^\emptyset \mid \drb{T}_X^M
               \mid \mcis \circ (\pdir_m \mid \drb{Q}_X^M))
  & \quad & n\in X,\ m\notin X \\
  \drb{\cell{\cofuse_n \mcomp S}{Q}}_X^M & =
  /m{:\,} \mathsf{h} \circ (\fdir_m \mid \mext \circ
  (\fmemb_{n,m} \circ \drb{S}_X^M
               \mid \mcis \circ \drb{Q}_X^{M \uplus \set{\fuse_n}}))
  & \quad & n\in X,\ m\notin X \\
  \drb{\cell{\cofuse_n \mcomp S}{Q}}_X^{M \uplus \set{\fuse_n}} & =
  /m{:\,} \mathsf{h} \circ \mext \circ (\fmemb_{n,m} \mid \drb{S}_X^M
               \mid \mcis \circ (\fdir_m \mid \drb{Q}_X^M))
  & \quad & n\in X,\ m\notin X \\
  \drb{K(\rho)}_X^M & = X \parallel
  (
   ( \textstyle
     \bigparallel_{i=1}^{s(K)} \dsb{\rho(i)}^{z_{i}}_X
   ) \circ
   K_{z_{1}\dots z_{s(K)}}
  )
  & \quad & K\in \set{A,P} \\
  \text{where} & \qquad 
  \dsb{h}_{X}^{z} = /z{:\,} \mathsf{h} \qquad
  \dsb{v}_{X}^{z} = /z{:\,} \mathsf{v} \qquad
  \dsb{x}_{X}^{z} = x / z & \quad & x \in X
\end{alignat*}

Notice that, for the fact that forgetting membranes (and hence
actions), calculus ??  corresponds exactly to the $\kappa$-calculus,
so in such case the previous encoding becomes the same of the one
shown in \cite{???} for encoding the $\kappa$-calculus into
biobigraphs\footnote{More precisely into protein link graphs, i.e.,
  biobigraphs restricted on the link graph part.}.

\begin{theorem}[Soundness]
  For every $\Gamma_{1}; \Gamma_{2} \vdash P : \tau$, if
  $\tau\restrict{\Gamma_{1}} = \emptyset$ then
  $\drb{P}_{fn(P)}^{\emptyset} \in \cat{BioBg}(\epsilon, \langle
  1,fn(P) \rangle)$.
\end{theorem}

\begin{theorem}[Completeness]
  For every bigraph $G \in \cat{BioBg}(\epsilon, \langle 1,X
  \rangle)$, there exists $P$ such that $\drb{P}_{X}^{\emptyset} =
  \hat{G}$ and $\Gamma_{1}, \Gamma_{2} \vdash P : \tau$ for some
  $\tau$, $\Gamma_{1}$, $\Gamma_{2}$ such that $\Gamma_{1} \cup
  \Gamma_{2} \subseteq X$.
\end{theorem}

\medskip
QUI METTERE LA CODIFICA DELLE REGOLE DI INTRODUZIONE E FAR VEDERE CHE
SODDISFANO LE CONDIZIONI IMPOSTE SUI BIOBIGRAFI.

\medskip
POI FAR VEDERE CHE LE REGOLE DI COMMITMENT SONO CASI SPECIALI DI
QUELLE DATE NEI BIOBIGRAFI.